\documentstyle[12pt]{article}
\newcommand{\bfp}{\mbox{\boldmath $p$}}

\newcommand{\bfh}{\mbox{\boldmath $h$}}
\newcommand{\bfq}{\mbox{\boldmath $q$}}
\def\ii{\'{\i}}
\def\nostrocostruttino#1\over#2{\mathrel{\mathop{\kern 0pt \rlap
{\hbox{$#1$}}} \hbox{\kern-.135em $#2$}}}

\newcommand{\ZP}[1]{{\it Z.\ Phys.}\ {\bf #1}}
\newcommand{\PL}[1]{{\it Phys.\ Lett.}\ {\bf #1}}
\newcommand{\PR}[1]{{\it Phys.\ Rev.}\ {\bf #1}}

\newcommand{\beq}{\begin{equation}}
\newcommand{\eeq}{\end{equation}}
\newcommand{\barr}{\begin{eqnarray}}
\newcommand{\earr}{\end{eqnarray}}
\newcommand{\ba}{\begin{array}}
\newcommand{\ea}{\end{array}}
\def\lsim{\mathrel{\rlap{\lower4pt\hbox{\hskip1pt$\sim$}}\raise1pt\hbox{$<$}}} 
\def\gsim{\mathrel{\rlap{\lower4pt\hbox{\hskip1pt$\sim$}}\raise1pt\hbox{$>$}}}
\newcommand{\ee}{e^-e^+}
\newcommand{\qq}{q\bar q}
\newcommand{\la}{\lambda}
\begin{document}
\begin{flushright}
DFTT 25/97 \\
INFNCA-TH9707 \\
hep-ph/9704420 \\
\end{flushright}
\vskip 1.5cm
\begin{center}
{\bf 
Off-diagonal helicity density matrix elements for vector mesons
produced at LEP
}\\
\vskip 1.5cm
{\sf M. Anselmino$^1$, M. Bertini$^{1,2}$, F. Murgia$^3$ and 
P. Quintairos$^{1,4}$}
\vskip 0.8cm
{$^1$Dipartimento di Fisica Teorica, Universit\`a di Torino and \\
      INFN, Sezione di Torino, Via P. Giuria 1, 10125 Torino, Italy\\
\vskip 0.5cm
$^2$Institut de Physique Nucl\'eaire de Lyon \\
43 Bvd. du 11 Novembre 1918, F-69622 Villeurbanne Cedex \\
\vskip 0.5cm
$^3$INFN, Sezione di Cagliari, Via A. Negri 18, 09127 Cagliari, Italy \\
\vskip 0.5cm
$^4$Centro Brasileiro de Pesquisas F\ii sicas \\
R. Dr. Xavier Sigaud 150, 22290-180, Rio de Janeiro, Brazil} \\

\end{center}
\vskip 1.5cm
\noindent
{\bf Abstract:}

\vspace{6pt}

\noindent
Final state $\qq$ interactions may give origin to non zero values of the 
off-diagonal element $\rho_{1,-1}$ of the helicity density matrix of vector 
mesons produced in $e^+ e^-$ annihilations, as confirmed by recent OPAL 
data on $\phi$ and $D^*$'s. Predictions are given for $\rho_{1,-1}$
of several mesons produced at large $z$ and small $p_T$, {\it i.e.}
collinear with the parent jet; the values obtained for $\phi$ and $D^*$ 
are in agreement with data.
\newpage
\pagestyle{plain}
\setcounter{page}{1}
{\bf 1 - Introduction}
\vskip 12pt
The spin properties of hadrons inclusively produced in high energy 
interactions are related to the fundamental properties of quarks and 
gluons and to their elementary interactions in a much more subtle way 
than unpolarized quantities; the usual hadronization models -- successful 
in predicting unpolarized cross-sections -- may not be adequate to 
describe spin effects, say the fragmentation of a polarized quark.

In Refs. \cite{akp} and \cite{aamr} it was pointed out how the
final state interactions between the $q$ and $\bar q$ produced
in $e^+ e^-$ annihilations -- usually neglected, but indeed 
necessary -- might give origin to non zero spin observables which would 
otherwise be forced to vanish. The off-diagonal matrix element $\rho_{1,-1}$ 
of vector mesons may be sizeably different from zero \cite{akp} due to a 
coherent fragmentation process which takes into account $q \bar q$ 
interactions; the incoherent fragmentation of a single independent quark 
leads to zero values for such off-diagonal elements. 
The same situation is not true for spin 1/2 baryons, for which the coherent 
fragmentation process only induces corrections which vanish in the limit
of small transverse momentum, $p_T$, of the quark inside the jet
\cite{aamr}. Both predictions, a non zero value of $\rho_{1,-1}$ for
$D^*$ and possibly $\phi$ particles \cite{opal}, and a value 
$\rho_{+-} \simeq (p_T/z\sqrt s)$ for $\Lambda$ ({\it i.e.}, its transverse 
polarization) \cite{aleph} have recently been confirmed experimentally.

We consider here in greater details the coherent fragmentation process
of $q\bar q$ produced at LEP, where the quarks are strongly polarized;
we are actually able to give predictions for $\rho_{1,-1}$ of several
vector mesons $V$ provided they are produced in two jet events, carry
a large momentum or energy fraction $z=2E_V/\sqrt s$, and have a small
transverse momentum $p_T$ inside the jet. Our estimates are in agreement
with the existing data and are crucially related both to the presence 
of final state interactions and to the Standard Model couplings of the 
elementary $e^- e^+ \to q \bar q$ interaction. 

In the next Section we review the formalism to compute the helicity
density matrix of a hadron produced in $e^- e^+ \to q \bar q \to h+X$
processes and give analytical expressions for the non diagonal matrix 
element $\rho_{1,-1}$ in case of final spin 1 hadrons; in Section 3
we obtain numerical estimates and in Section 4 we make some further 
comments and conclusions.

\vskip 12pt
\noindent
{\bf 2 -} {\mbox{\boldmath $\rho^{\,}_{1,-1}(V)$}} {\bf in the process} 
{\mbox{\boldmath $e^- e^+ \to q\bar q \to V + X$}}
\vskip 12pt
The helicity density matrix of a hadron $h$ inclusively produced in the 
two jet event $e^- e^+ \to q\bar q \to h + X$ can be written as 
\cite{akp, aamr}
\beq
\rho^{\,}_{\la^{\,}_h \la^{\prime}_h}(h) 
= {1\over N_h} \sum_{q,X,\la^{\,}_X,\la^{\,}_q,\la^{\,}_{\bar q},
\la^{\prime}_q,\la^{\prime}_{\bar q}} 
D^{\,}_{\la^{\,}_h \la^{\,}_X; \la^{\,}_q,\la^{\,}_{\bar q}} \>\>
\rho^{\,}_{\la^{\,}_q,\la^{\,}_{\bar q};
\la^{\prime}_q,\la^{\prime}_{\bar q}}\,(\qq) \>\> 
D^*_{\la^{\prime}_h \la^{\,}_X; \la^{\prime}_q,\la^{\prime}_{\bar q}} \,,
\label{rhoh}
\eeq
where $\rho^{\,}_{\la^{\,}_q,\la^{\,}_{\bar q};\la^{\prime}_q,
\la^{\prime}_{\bar q}}\,(\qq)$ is the helicity density matrix of the 
$q\bar q$ state created in the annihilation of the unpolarized 
$e^+$ and $e^-$,
\beq
\rho^{\,}_{\la^{\,}_q,\la^{\,}_{\bar q};
\la^{\prime}_q,\la^{\prime}_{\bar q}}\,(\qq)
= {1\over 4N_{\qq}} \sum_{\la^{\,}_{-}, \la^{\,}_{+}}
M^{\,}_{\la^{\,}_q \la^{\,}_{\bar q};\la^{\,}_{-} \la^{\,}_{+}} \>
M^*_{\la^{\prime}_q \la^{\prime}_{\bar q}; \la^{\,}_{-} \la^{\,}_{+}} \,.
\label{rhoqq}
\eeq
The $M$'s are the helicity amplitudes for the $\ee \to \qq$ process and
the $D$'s are the fragmentation amplitudes, {\it i.e.} the helicity
amplitudes for the process $\qq \to h+X$; the $\sum_{X,\lambda_X}$ stands 
for the phase space integration and the sum over spins of all the unobserved 
particles, grouped into a state $X$. The normalization factors $N_h$ and
$N_{\qq}$ are given by
\beq
N_h = \sum_{q,X; \la^{\,}_h, \la^{\,}_X, \la^{\,}_q, \la^{\,}_{\bar q},
\la^{\prime}_q, \la^{\prime}_{\bar q}} 
D^{\,}_{\la^{\,}_h \la^{\,}_X; \la^{\,}_q,\la^{\,}_{\bar q}} \>\>
\rho^{\,}_{\la^{\,}_q,\la^{\,}_{\bar q};
\la^{\prime}_q,\la^{\prime}_{\bar q}}\,(\qq) \>\> 
D^*_{\la^{\,}_h \la^{\,}_X; \la^{\prime}_q,\la^{\prime}_{\bar q}} \,
= \sum_q D^h_q \,,
\label{nh}
\eeq
where $D^h_q$ is the usual fragmentation function of quark $q$ into 
hadron $h$ [see also comment after Eq. (\ref{dhq})], and 
\beq
N_{\qq} = {1\over 4} 
\sum_{\la^{\,}_q, \la^{\,}_{\bar q}; \la^{\,}_{-}, \la^{\,}_{+}} \vert 
M^{\,}_{\la^{\,}_q \la^{\,}_{\bar q}; \la^{\,}_{-} \la^{\,}_{+}} \vert^2 \,.
\label{nqq}
\eeq

The center of mass helicity amplitudes for the $\ee \to \qq$ process can 
be computed in the Standard Model and are given by
\barr
& & M_{\la^{\,}_q \la^{\,}_{\bar q};\la^{\,}_{-} \la^{\,}_{+}}(s,\theta) 
= \ e^2\  \delta_{\la^{\,}_{-}, -\la^{\,}_{+}} \>
\delta_{\la^{\,}_q, -\la^{\,}_{\bar q}} \times \nonumber \\
&\times& \Biggl\{ \biggl[ e_q  - 
g_{_Z}(s) \, g_{_V}^l \, g_{_V}^q \,
\biggr] (1 + 4 \la^{\,}_{-} \la^{\,}_q \cos\theta) \nonumber \\
&+& g_{_Z}(s)  
\biggl[ 2\,g_{_V}^l\,g_{_A}^q(\la^{\,}_{-}\cos\theta + \la^{\,}_q) 
\label{msm} \\
&+& 2\,g_{_A}^l\,g_{_V}^q
(\la^{\,}_{-} + \la^{\,}_q \cos\theta) - g_{_A}^l\,g_{_A}^q(\cos\theta +
4\la^{\,}_{-}\la^{\,}_q) \biggr] \Biggr\} \,, \nonumber 
\earr
where $\sqrt s$ is the total $e^+ e^-$ c.m. energy, $\theta$ the $q$ 
production angle ({\it i.e.} the angle between the incoming $e^-$ and 
the outgoing $q$) and $e_q$ is the quark charge. Lepton and quark masses 
have been neglected with respect to their energies and we report here
for convenience the Standard Model coupling constants: 
\barr
g_{_V}^l &=& -{1\over 2} + 2\sin^2\theta_{_W} \quad\quad g_{_A}^l = -{1\over 2} 
\nonumber \\
g_{_V}^{u,c,t} &=&  \>\> {1\over 2} - {4\over 3}\sin^2\theta_{_W} \quad\quad
g_{_A}^{u,c,t} = {1\over 2} \label{cc} \\
g_{_V}^{d,s,b} &=& -{1\over 2} + {2\over 3}\sin^2\theta_{_W} \quad\quad
g_{_A}^{d,s,b} = -{1\over 2}  \nonumber \\
g_{_Z}(s) &=& \frac{1}{4 \sin^2 \theta_{_W} \cos^2 \theta_{_W}}\  
              \frac{s}{(s-M^2_{_Z}) + i M_{_Z} \Gamma_{_Z}} \,\cdot
\nonumber
\earr 

From Eqs. (\ref{rhoqq}), (\ref{nqq}) and (\ref{msm}) one finds the explicit
expressions of the only non zero elements of $\rho(\qq)$:
 
\begin{eqnarray}
\rho^{\,}_{+-;+-}(\qq) &=& 1- \rho^{\,}_{-+;-+}(\qq) \>=\>  
\frac{a'_q (s) \ (1+\cos^2 \theta) - b'_q(s)  \cos \theta}   
{\mu_q(s)\ (1+\cos^2 \theta) + \eta_q(s) \cos \theta} \label{rhoqqan++} \\ 
\rho^{\,}_{+-;-+}(\qq) &=& \rho^{*\,}_{-+;+-}(\qq) \>=\>   
\frac{ \left[a_q (s) - i b_q(s) \right]\ \sin^2 \theta }   
{\mu_q(s)\ (1+\cos^2 \theta) + \eta_q(s) \cos \theta} 
\label{rhoqqan+-} 
\end{eqnarray}
where $+,-$ stand for helicity $+1/2$ and $-1/2$ and  
where, for an arbitrary total energy $\sqrt{s}$,  
\begin{eqnarray} 
a'_q (s) &=&  e_q^2 + |g_{_Z}(s)|^2 \ (g_{_V} - g_{_A})_{q}^2 \ 
(g_{_V}^2 + g_{_A}^2)_{l}
-2e_q \ {\rm Re} [g_{_Z}(s)] \ g_{_V}^l (g_{_V} - g_{_A})_{q}  \nonumber \\
b'_q (s) &=& 4g_{_A}^l (g_{_V} -g_{_A})_{q} \ \left[\ |g_{_Z}(s)|^2 \ 
g_{_V}^l (g_{_V} - g_{_A})_{q} - e_q \ {\rm Re} [g_{_Z}(s)] \right] 
\nonumber \\
\nonumber \\
a_q (s) &=&  e_q^2 +|g_{_Z}(s)|^2\ (g^2_{_V} - g^2_{_A})_{q} \ 
(g^2_{_V} + g^2_{_A})_{l} -2e_q \ g^l_{_V} g^q_{_V} \ {\rm Re} 
[g_{_Z}(s)] \label{ab} \\
b_q(s) &=& - 2e_q \ g^l_{_V}  g^q_{_A} \ {\rm Im} [g_{_Z}(s)]  \nonumber \\ 
\nonumber \\
\mu_q(s) &=& 2\left[\ e_q^2 +|g_{_Z}(s)|^2 \ (g^2_{_V} + g^2_{_A})_{l} 
(g^2_{_V} + g^2_{_A})_{q}
-2e_q\ g^l_{_V} g^q_{_V}\ {\rm Re} [g_{_Z}(s)]  \right]  \nonumber \\ 
\eta_q(s) &=&  8g^l_{_A} g^q_{_A} \ \left[\ 2 \ |g_{_Z}(s)|^2 \ 
g^l_{_V} g^q_{_V} - e_q \ {\rm Re} [g_{_Z}(s)] \right] \, \nonumber 
\end{eqnarray}

\vspace{0.5cm}
\noindent which at $\sqrt s = M_{_Z}$ read
 
\barr 
a'_q &=&  e^2_q + \zeta^2 \, (g_{_V} - g_{_A})_q^2 \ (g^2_{_V} + g^2_{_A})_{l} 
\nonumber \\
b'_q &=&  4 \zeta^2 \, (g_{_A} g_{_V})_{l} \ (g_{_V} - g_{_A})_q^2 \nonumber \\
a_q &=& e_q^2 + \zeta^2 \,(g^2_{_V} - g^2_{_A})_{q} \ (g^2_{_V} + g^2_{_A})_{l} 
\nonumber \\ 
b_q &=& 2 e_q \ \zeta \, g^l_{_V}  g^q_{_A} \label{eq5} \\ 
\mu_q &=&2 [\ e_q^2+\zeta^2\ (g^2_{_V} + g^2_{_A})_{l} \ (g^2_{_V} + 
g^2_{_A})_{q}] \nonumber \\ 
\eta_q &=& 16 \zeta^2 \, (g_{_A} g_{_V})_{l} \ (g_{_A} g_{_V})_{q} \nonumber\\
\zeta &=& {M_{_Z} \over 4 \ \Gamma_{_Z} \, \sin^2\theta_{_W} \, 
\cos^2\theta_{_W}} \,\cdot \nonumber
\earr

Eqs. (\ref{rhoqqan++}) and (\ref{rhoqqan+-}) hold for the production 
of a quark with flavour $q$ at a c.m. angle $\theta$, defined as the angle 
between the incoming negative lepton and the outgoing quark; in the 
$p_T \to 0$ limit this is the same angle as the production angle of the 
observed hadron $h$. However, 
$h$ can be produced also in the fragmentation of an antiquark $\bar q$ and 
the $\sum_q$ in Eqs. (\ref{rhoh}) and (\ref{nh}) takes into account also
this possibility ($q = u, d, s, c, b, \bar u, \bar d, \bar s, \bar c, \bar b$):
the helicity density matrix $\rho(\bar qq)$ for the production of an
antiquark at the angle $\theta$ can be obtained from $\rho(\qq)$ with
the simple replacements:
\barr
\rho^{\,}_{+-;+-}(\bar qq, \theta) &=& 
\rho^{\,}_{-+;-+}(\qq, \pi - \theta) \nonumber \\ 
\rho^{\,}_{+-;-+}(\bar qq, \theta) &=& 
\rho^*_{+-;-+}(\qq, \pi - \theta) \,.
\label{ant}
\earr

The expressions (\ref{rhoqqan+-}), (\ref{ab}) and (\ref{eq5}) are exact and 
contain both electromagnetic and weak interaction contributions. However, at 
LEP energy $\sqrt s = M_{_Z}$, the weak contribution dominates, $\zeta \gg 1$
in Eqs. (\ref{eq5}); if one also takes into account that $\eta_q$ is 
depressed by the small value of $g^l_{_V}$ a simple approximate and useful 
formula for $\rho^{\,}_{+-;-+}$ is given by [for an exact value at 
$\sqrt s = M_{_Z}$ see Eqs. (\ref{rho+--+})]
\beq
\rho^{Z}_{+-;-+}(\qq) \simeq {1\over 2}\,{(g^2_{_V} - g^2_{_A})_q \over 
(g^2_{_V} + g^2_{_A})_q} \, {\sin^2\theta \over 1+ \cos^2\theta} \, \cdot
\label{rhoqqap}
\eeq
Eq. (\ref{rhoqqap}) clearly shows the $\theta$ dependence of
$\rho^{\,}_{+-;-+}$. This approximate expression is the same
both for $\rho(\qq)$ and $\rho(\bar qq)$. In the case of pure
electromagnetic interactions ($\sqrt s \ll M_{_Z}$) one has exactly:
\beq
\rho^{\gamma}_{+-;-+}(\qq) = {1\over 2}
\,{\sin^2\theta \over 1+ \cos^2\theta} \, \cdot
\label{rhoqqelm}
\eeq
Notice that Eqs. (\ref{rhoqqap}) and (\ref{rhoqqelm}) have the same
angular dependence, but a different sign for the coefficient in front,
which is negative for the $Z$ contribution [see Eqs. (\ref{cc})]. 

By using the above equations for $\rho(\qq)$ [and $\rho(\bar qq)$] into 
Eq. (\ref{rhoh}) one obtains the most general expression of $\rho(h)$ in 
terms of the $\qq$ spin state and the unknown fragmentation amplitudes 
\cite{aamr}. Such expression can be greatly simplified if one considers the 
production of hadrons almost collinear with the parent jet: the 
$q \bar q \to h + X$ fragmentation is then essentially a c.m. forward 
process and the unknown $D$ amplitudes must satisfy the angular momentum 
conservation relation \cite{bls}
\beq
D^{\,}_{\la^{\,}_h \la^{\,}_X; \la^{\,}_q,\la^{\,}_{\bar q}} \>
D^*_{\la^{\prime}_h \la^{\,}_X; \la^{\prime}_q,\la^{\prime}_{\bar q}}
\sim \left( \sin{\theta_h\over 2}
\right)^{
\vert \la^{\,}_h - \la^{\,}_X - \la^{\,}_q + \la^{\,}_{\bar q} \vert
+ \vert \la^{\prime}_h - \la^{\,}_X - \la^{\prime}_q + \la^{\prime}_{\bar q} 
\vert} \,,
\label{frd}
\eeq
where $\theta_h$ is the angle between the hadron momentum, 
$\bfh = z \bfq + \bfp_T$, and the quark momentum $\bfq$, that is
\beq \sin\theta_h \simeq {2p_T \over z\sqrt s} \,\cdot \label{ptb} \eeq
The bilinear combinations of fragmentation amplitudes contributing to
$\rho(h)$ are then not suppressed by powers of $(p_T/(z\sqrt s))$ only
if the exponents in Eq. (\ref{frd}) are zero; which yields 
\beq
\la^{\,}_X = \la^{\,}_h - (\la^{\,}_q - \la^{\,}_{\bar q}) =
\la^{\prime}_h - (\la^{\prime}_q - \la^{\prime}_{\bar q})\,. 
\label{las}
\eeq

In the $p_T \to 0$ limit one has then the simple result for the non diagonal 
density matrix elements of spin 1 mesons \cite{akp, aamr}:
\barr
{\rm Re}[\rho_{1,-1}(V)] &=& {1\over N_h} \sum_{X,q}
D_{10;+-} \> D^*_{-10;-+} \> {\rm Re}[\rho_{+-;-+}(\qq)] 
\label{rho1-1r} \\
{\rm Im}[\rho_{1,-1}(V)] &=& {1\over N_h} \sum_{X,q}
D_{10;+-} \> D^*_{-10;-+} \> {\rm Im}[\rho_{+-;-+}(\qq)] 
\label{rho1-1i}  
\earr
with
\beq
N_h = \sum _q D^h_q = \sum_{q,X; \la^{\,}_h, \la^{\,}_X} \Bigl[
\vert D^{\,}_{\la^{\,}_h \la^{\,}_X; +-}\vert^2 \>
\rho^{\,}_{+-;+-}(\qq) + \vert D^{\,}_{\la^{\,}_h \la^{\,}_X; -+} \vert^2
\> \rho^{\,}_{-+;-+}(\qq) \Bigr] \,.
\label{nh0}
\eeq

Eq. (\ref{rho1-1r}) and (\ref{rho1-1i}) explicitely show that the coherent 
quark fragmentation allows non zero off-diagonal helicity density matrix 
elements which, for vector mesons, survive also in the small $p_T$ limit; 
the other off-diagonal matrix elements for spin 1 particles and all 
off-diagonal matrix elements for spin 1/2 particles are bound, via 
Eq. (\ref{frd}), to vanish at small $p_T/\sqrt s$ values \cite{akp, aamr}. 
Recent experimental data have confirmed both the non zero value of 
$\rho_{1,-1}(D^*)$ \cite{opal} and the small value of 
$\rho_{+-}(\Lambda)$ \cite{aleph}.

In the next Section we give numerical estimates of $\rho_{1,-1}$ for
several vector mesons, exploiting Eq. (\ref{rho1-1r}) and the fact that,
at least for valence quark contributions, the dependence on the 
fragmentation amplitudes either cancels out or can be expressed in 
terms of other measured quantities. 

\vskip 12pt
\noindent
{\bf 3 - Numerical estimates of} {\mbox{\boldmath $\rho^{\,}_{1,-1}(V)$}} 
{\bf at} {\mbox{\boldmath $\sqrt s = M_{_Z}$}}
\vskip 12pt

Let us consider Eqs. (\ref{rho1-1r})-(\ref{nh0}).
Despite our ignorance of the fragmentation amplitudes we see that 
in the $p_T \to 0$ limit, due to Eqs. (\ref{frd}) and (\ref{las}), 
only few of them give a leading contribution; moreover,
the fragmentation is a parity conserving forward process, so that
the fragmentation amplitudes must satisfy the relationship \cite{bls}
\beq
D^{\,}_{-\la^{\,}_h -\la^{\,}_X; -+} = (-1)^{S^{\,}_h + S^{\,}_X + 
\la^{\,}_h - \la^{\,}_X} \> 
D^{\,}_{\la^{\,}_h \la^{\,}_X; +-} \,,
\label{par}
\eeq
where $S^{\,}_h$ and $S^{\,}_X$ are respectively the spin of hadron $h$ 
and of the unobserved system $X$ (the intrinsic parities of the initial 
and final states must be the same). In particular Eq. (\ref{par}) for 
spin 1 hadrons yields 
\beq
D^{\,}_{-10;-+} = (-1)^{S^{\,}_X} \> D^{\,}_{10; +-} \,.
\label{par0}
\eeq

Notice that the parity relationship (\ref{par}) and Eq. (\ref{rhoqqan++})
allow to write:
\beq
D^h_q = \sum_{X; \la^{\,}_h, \la^{\,}_X}
\vert D^{\,}_{\la^{\,}_h \la^{\,}_X; +-} \vert^2 \,,
\label{dhq}
\eeq
which is the fragmentation function of quark $q$ into hadron $h$, whose
spin is not observed; such fragmentation function is independent of the
quark polarization, described by $\rho(\qq)$. Instead, the fragmentation
functions of a polarized quark $q$ into a hadron $h$ with helicity
$\la^{\,}_h$ are given by:
\barr
D^{h,\la^{\,}_h}_q &=& \sum_{X; \la^{\,}_X} \Bigl[
\vert D^{\,}_{\la^{\,}_h \la^{\,}_X; +-} \vert^2 \> \rho_{+-;+-}(\qq)
+ \vert D^{\,}_{\la^{\,}_h \la^{\,}_X; -+} \vert^2 \> \rho_{-+;-+}(\qq) \Bigl] 
\nonumber \\
&=& D^{h,\la^{\,}_h}_{q,+} \> \rho_{+-;+-}(\qq)
+ D^{h,\la^{\,}_h}_{q,-} \> \rho_{-+;-+}(\qq) \,,
\label{dhql}
\earr
which is consistent with $\sum_{\la^{\,}_h} D^{h,\la^{\,}_h}_q = D^h_q$ and 
where $D^{h,\la^{\,}_h}_{q,\la^{\,}_q}$ is the fragmentation function of quark 
$q$ with helicity $\la^{\,}_q$ into hadron $h$ with helicity $\la^{\,}_h$; 
$\rho^{\,}_{\la^{\,}_q, -\la^{\,}_q; \la^{\,}_q, -\la^{\,}_q}\,(\qq)$ is
the probability for $q$ to have helicity $\la^{\,}_q$.

Taking into account Eq. (\ref{las}) and (\ref{par}) the above fragmentation 
functions read:
\barr
D^h_q &=& \sum_X \Bigl[ \vert D^{\,}_{10;+-} \vert^2 + \vert D^{\,}_{0-1;+-} 
\vert^2 + \vert D^{\,}_{-1-2;+-} \vert^2 \Bigr] \nonumber \\
&=& D^{h,1}_{q,+} + D^{h,0}_{q,+} + D^{h,-1}_{q,+} \label{dqhu} \\
D^{h,1}_q &=& \sum_X \Bigl[ \vert D^{\,}_{10;+-} \vert^2 \> \rho_{+-;+-}(\qq)
+ \vert D^{\,}_{12; -+} \vert^2 \> \rho_{-+;-+}(\qq) \Bigr] 
\nonumber \\
&=& D^{h,1}_{q,+} \> \rho_{+-;+-}(\qq) + D^{h,1}_{q,-} \> \rho_{-+;-+}(\qq)
\label{dqh1} \\
D^{h,0}_q &=& \sum_X \vert D^{\,}_{0-1;+-} \vert^2 
= D^{h,0}_{q,+} \label{dqh0} \\
D^{h,-1}_q &=& \sum_X \Bigl[ \vert D^{\,}_{12;-+} \vert^2 \> \rho_{+-;+-}(\qq)
+ \vert D^{\,}_{10; +-} \vert^2 \> \rho_{-+;-+}(\qq) \Bigr] 
\nonumber \\
&=& D^{h,1}_{q,-} \> \rho_{+-;+-}(\qq) + D^{h,1}_{q,+} \> \rho_{-+;-+}(\qq)
\label{dqh-1}
\,.
\earr

We now assume that, at least for valence quarks:
\barr
D^{h,1}_{q,-} &=& D^{h,-1}_{q,+} = 0 \label{ass1} \\
D^{h,0}_{q,+} &=& \alpha^V_q \> D^{h,1}_{q,+} \label{ass2} \,.
\earr
The first of these assumptions simply means that quarks with helicity
1/2 ($-1/2$) cannot fragment into vector mesons with helicity $-1$ ($+1$).
This is true for valence quarks assuming vector meson wave functions 
with no orbital angular momentum, like in $SU(6)$. The second assumption 
is also true in $SU(6)$ with $\alpha^V_q = 1/2$ for 
any valence $q$ and $V$. Rather than taking  
$\alpha^V_q = 1/2$ we prefer to relate the value of $\alpha^V_q$ to the 
value of $\rho^{\,}_{0,0}(V)$ which can be or has been measured.
In fact, always in the $p_T \to 0$ limit, one has, from Eqs. (\ref{rhoh}),  
(\ref{las}), (\ref{par}), (\ref{ass1}) and (\ref{ass2}):
\beq
\rho^{\,}_{0,0}(V) = {\sum_q \alpha^V_q \, D^{h,1}_{q,+}
\over \sum_q \> (1+\alpha^V_q) \, D^{h,1}_{q,+}} \,\cdot
\label{rho00}
\eeq
If $\alpha^V_q$ is the same for all valence quarks in $V$ 
($\alpha^V_q = \alpha^V$) 
one has, for the valence quark contribution:
\beq
\alpha^V = {\rho^{\,}_{0,0}(V) \over 1 - \rho^{\,}_{0,0}(V)} \,\cdot
\label{alrho}
\eeq
Notice that the $SU(6)$ value $\alpha^V_q = 1/2$ correspond to 
$\rho^{\,}_{0,0} = 1/3$, that is no alignment, 
$A= (1/2)(3\rho^{\,}_{0,0}-1) = 0$, for the vector meson. 

If we now use Eqs. (\ref{par0}), (\ref{dqhu}), (\ref{ass1}), 
(\ref{ass2}) into Eqs. (\ref{rho1-1r}) and (\ref{rho1-1i}) we obtain
\beq
\rho^{\,}_{1,-1}(V) = {\sum_{q,X} \> (-1)^{S_{_X}} \, \vert D^{\,}_{10;+-} 
\vert^2 \> \rho_{+-;-+}(\qq) \over \sum_{q,X} \> (1 + \alpha^V_q) \,  
\vert D^{\,}_{10;+-} \vert^2} \,\cdot
\label{rho1-1t}
\eeq

The numerator in the above equation depends on the squared amplitude 
$\vert D^{\,}_{10;+-} \vert^2$ for the $\qq \to V+X$ forward fragmentation
process and on $S_{_X}$. The $\qq$ state is such that $J=J_z=1$; the final 
undetected system $X$ must then have $\la^{\,}_X=0$ with $S_{_X}=0,1$ or 2,  
the only states which can combine with the $S_h=\la^{\,}_h=1$ vector meson 
state to give a $VX$ spin state with $J=J_z=1$. On a simple statistical 
basis these 3 possible states have respectively relative probabilities 
1, 1/6 and 1/30. One can then conclude that the $S_{_X}=0$ state dominates 
and approximate the above equation (\ref{rho1-1t}) with
\beq
\rho^{\,}_{1,-1}(V) \simeq {\sum_q \, D^{V,1}_{q,+} \> \rho_{+-;-+}(\qq) 
\over \sum_q \> (1 + \alpha^V_q) \> D^{V,1}_{q,+}} \,\cdot
\label{rho1-1ts}
\eeq
The actual value (\ref{rho1-1t}) should only be slightly smaller, due
to some contribution from $S_{_X}=1$.

Again, if only one flavour contributes or if we can assume that
$\alpha^V_q$ does not depend on the valence quark flavour, Eq. (\ref{alrho})
further simplifies Eq. (\ref{rho1-1ts}) to
\beq
\rho^{\,}_{1,-1}(V) \simeq [1 - \rho^{\,}_{0,0}(V)] \,
{\sum_q \, D^{V,1}_{q,+} \> \rho_{+-;-+}(\qq) 
\over \sum_q \, D^{V,1}_{q,+}} \,\cdot
\label{rho1-1tss}
\eeq
We shall now consider some specific cases in which we expect 
Eq. (\ref{rho1-1tss}) to hold; let us remind once more that our 
conclusions apply to spin 1 vector mesons produced in 
$e^- e^+ \to q \bar q \to V+X$ processes in the limit of small $p_T$
and large $z$, {\it i.e.}, to vector mesons produced in two jet events
($e^- e^+ \to \qq$) and collinear with one of them ($p_T = 0$), 
which is the jet generated by a quark which is a valence quark for the
observed vector meson (large $z$). These conditions should be met 
more easily in the production of heavy vector mesons. 

Let us then start from the cases $V=B^{*\pm,0},\> D^{*\pm,0}$. In such 
a case one can safely assume that the fragmenting quark is 
the heavy one so that Eq. (\ref{rho1-1tss}) applies and one has:
\barr
\rho^{\,}_{1,-1}(B^{*+}) &\simeq& [1 - \rho^{\,}_{0,0}(B^{*+})] \>
\rho_{+-;-+}(\bar bb) \nonumber \\
\rho^{\,}_{1,-1}(B^{*-}) &\simeq& [1 - \rho^{\,}_{0,0}(B^{*-})] \>
\rho_{+-;-+}(b\bar b) \label{bstar} \\
\rho^{\,}_{1,-1}(B^{*0}) &\simeq& [1 - \rho^{\,}_{0,0}(B^{*0})] \>
\rho_{+-;-+}(\bar bb) \nonumber \\
\nonumber \\
\rho^{\,}_{1,-1}(D^{*+}) &\simeq& [1 - \rho^{\,}_{0,0}(D^{*+})] \>
\rho_{+-;-+}(c\bar c) \nonumber \\
\rho^{\,}_{1,-1}(D^{*-}) &\simeq& [1 - \rho^{\,}_{0,0}(D^{*-})] \>
\rho_{+-;-+}(\bar cc) \label{dstar} \\
\rho^{\,}_{1,-1}(D^{*0}) &\simeq& [1 - \rho^{\,}_{0,0}(D^{*0})] \>
\rho_{+-;-+}(c\bar c) \nonumber 
\earr

Similarly one obtains:
\beq
\rho^{\,}_{1,-1}(\phi) \simeq {1\over 2} \> [1 - \rho^{\,}_{0,0}(\phi)] \>
[\rho_{+-;-+}(s\bar s) + \rho_{+-;-+}(\bar ss)]
\label{phi}
\eeq
where we have assumed $D^{\phi,1}_{s,+} = D^{\phi,1}_{\bar s,+}$,
as it should be.

For $\rho$'s, assuming all valence quark fragmentation functions 
to be the same, one has 
\barr
\rho^{\,}_{1,-1}(\rho^+) &\simeq& {1\over 2} \> [1 - \rho^{\,}_{0,0}(\rho^+)] 
\> [\rho_{+-;-+}(u\bar u) + \rho_{+-;-+}(\bar dd)] \nonumber \\
\rho^{\,}_{1,-1}(\rho^0) &\simeq& {1\over 4} \> [1 - \rho^{\,}_{0,0}(\rho^0)] 
\> [\rho_{+-;-+}(u\bar u) + \rho_{+-;-+}(d\bar d) \nonumber \\
&+& \rho_{+-;-+}(\bar uu) + \rho_{+-;-+}(\bar dd)]
\label{rho} \\
\rho^{\,}_{1,-1}(\rho^-) &\simeq& {1\over 2} \> [1 - \rho^{\,}_{0,0}(\rho^-)] 
\> [\rho_{+-;-+}(d\bar d) + \rho_{+-;-+}(\bar uu)] \,. \nonumber
\earr

The assumption that all valence quark fragmentation functions are the 
same is very natural for $\rho$'s, but it might be weaker for $K^*$ mesons; 
if nevertheless we assume that, at least at large $z$, 
$D^{K^{*+},1}_{\bar s,+} = D^{K^{*+},1}_{u,+}$, and similarly for 
$K^{*0}$ and $K^{*-}$ , we have
\barr
\rho^{\,}_{1,-1}(K^{*+}) &\simeq& {1\over 2} \> [1 - \rho^{\,}_{0,0}(K^{*+})] 
\> [\rho_{+-;-+}(u\bar u) + \rho_{+-;-+}(\bar ss)] \nonumber \\
\rho^{\,}_{1,-1}(K^{*0}) &\simeq& {1\over 2} \> [1 - \rho^{\,}_{0,0}(K^{*0})] 
\> [\rho_{+-;-+}(d\bar d) + \rho_{+-;-+}(\bar ss)] \label{kstar} \\
\rho^{\,}_{1,-1}(K^{*-}) &\simeq& {1\over 2} \> [1 - \rho^{\,}_{0,0}(K^{*-})] 
\> [\rho_{+-;-+}(\bar uu) + \rho_{+-;-+}(s \bar s)] \,. \nonumber
\earr
A predominant contribution of the $s$ quark would instead lead to results 
similar to those found for $B^*$.

Eqs. (\ref{bstar})-(\ref{kstar}) show how the value of $\rho^{\,}_{1,-1}(V)$
are simply related to the off-diagonal matrix element $\rho_{+-;-+}(\qq)$
of the $\qq$ pair created in the elementary $e^- e^+ \to \qq$ process;
such off-diagonal elements would not appear in the incoherent independent 
fragmentation of a single quark, yielding $\rho^{\,}_{1,-1}(V)=0$.

We can now make numerical predictions by inserting into the above equations 
the explicit values of $\rho_{+-;-+}(\qq)$ at $\sqrt s = M_{_Z}$, 
Eqs. (\ref{rhoqqan+-}), (\ref{eq5}) and (\ref{cc}) with 
$\sin^2\theta_{_W} = 0.2237$, $M_{_Z} = 91.19$ GeV, $\Gamma_{_Z} = 2.50$ GeV
\cite{pdb}:
\barr
\rho_{+-;-+}(u\bar u) &=& -0.36 \, (1 - 0.013\,i) \>
{\sin^2\theta \over (1 + \cos^2\theta) + 0.29 \cos\theta} \nonumber \\
\rho_{+-;-+}(\bar uu) &=& -0.36 \, (1 + 0.013\,i) \>
{\sin^2\theta \over (1 + \cos^2\theta) - 0.29 \cos\theta} \nonumber \\
\rho_{+-;-+}(d\bar d) &=& -0.17 \, (1 - 0.010\,i) \>
{\sin^2\theta \over (1 + \cos^2\theta) + 0.39 \cos\theta} \label{rho+--+} \\
\rho_{+-;-+}(\bar dd) &=& -0.17 \, (1 + 0.010\,i) \>
{\sin^2\theta \over (1 + \cos^2\theta) - 0.39 \cos\theta} \, \cdot \nonumber 
\earr
The values for $s,b$ and $c$ quarks are respectively the same as for
$d$ and $u$.

If we instead use for simplicity the approximate expressions (\ref{rhoqqap}), 
valid at $\sqrt s = M_{_Z}$ and which are the same for $\rho_{+-;-+}(\qq)$ and 
$\rho_{+-;-+}(\bar qq)$, we have the simple results
\barr
\rho^{\,}_{1,-1}(B^{*\pm,0}) &\simeq& -0.170 \ [1- \rho^{\,}_{0,0}(B^*)] 
\ {\sin^2\theta \over 1 + \cos^2\theta} \nonumber \\
&=& -(0.109 \pm 0.015) \ {\sin^2\theta \over 1 + \cos^2\theta} \\
\rho^{\,}_{1,-1}(D^{*\pm,0}) &\simeq& -0.360 \ [1- \rho^{\,}_{0,0}(D^*)] 
\ {\sin^2\theta \over 1 + \cos^2\theta} \nonumber \\
&=& -(0.216 \pm 0.007) \ {\sin^2\theta \over 1 + \cos^2\theta} \\
\rho^{\,}_{1,-1}(\phi) &\simeq& -0.170 \ [1- \rho^{\,}_{0,0}(\phi)] 
\ {\sin^2\theta \over 1 + \cos^2\theta} \nonumber \\
&=& -(0.078 \pm 0.014) \ {\sin^2\theta \over 1 + \cos^2\theta} \\
\rho^{\,}_{1,-1}(\rho^{\pm,0}) &\simeq& -0.265 \ 
[1- \rho^{\,}_{0,0}(\rho)] \ {\sin^2\theta \over 1 + \cos^2\theta} \\
\rho^{\,}_{1,-1}(K^{*\pm}) &\simeq& -0.265 \ 
[1- \rho^{\,}_{0,0}(K^*)] \ {\sin^2\theta \over 1 + \cos^2\theta} \\
\rho^{\,}_{1,-1}(K^{*0}) &\simeq& -0.170 \ 
[1- \rho^{\,}_{0,0}(K^*)] \ {\sin^2\theta \over 1 + \cos^2\theta} 
\earr

\vspace{0.5cm}
\noindent
where we have used $\rho^{\,}_{0,0}(B^{*\pm,0}) = 0.36 \pm 0.09$, 
$\rho^{\,}_{0,0}(D^{*\pm,0}) = 0.40 \pm 0.02$ and 
$\rho^{\,}_{0,0}(\phi) = 0.54 \pm 0.08$ \cite{opal}; no data are available
on $\rho^{\,}_{0,0}(\rho)$ and $\rho^{\,}_{0,0}(K^*)$. Notice that in such 
approximation $\rho^{\,}_{1,-1}(V)$ is real and that the $\cos\theta$ term 
in the denominator of Eq. (\ref{rhoqqan+-}) has been neglected. This term
would induce small differences between the values of $\rho^{\,}_{1,-1}(B^{*+})$
[or $\rho^{\,}_{1,-1}(D^{*+})$] and $\rho^{\,}_{1,-1}(B^{*-})$
[or $\rho^{\,}_{1,-1}(D^{*-})$]; it has much smaller effects on the values 
of $\rho^{\,}_{1,-1}(\phi)$, $\rho^{\,}_{1,-1}(\rho)$ and 
$\rho^{\,}_{1,-1}(K^*)$.  

Finally, in case one collects all meson produced at different angles in
the full available $\theta$ range (say $\alpha < \theta < \pi -\alpha, 
\> |\cos\theta| < \cos\alpha$) an average should be taken in $\theta$, 
weighting the different values of $\rho^{\,}_{1,-1}(\theta)$ with the 
cross-section for the $e^-e^+ \to V+X$ process; this amounts essentially 
to weight the values of $\rho_{+-;-+}(\qq; \theta)$ appearing in 
Eqs. (\ref{bstar})-(\ref{kstar}) and given in Eq. (\ref{rho+--+})
or (\ref{rhoqqap}) with the cross-section for the $e^-e^+ \to \qq$ process, 
proportional to the normalization factor $N_{\qq}$ given in Eq. (\ref{nqq}).
Such an average has a simple analytical expression if one uses the
approximate value (\ref{rhoqqap}):
\barr
\langle \rho^{\,}_{1,-1}(B^{*\pm,0}) \rangle_{[\alpha, \pi-\alpha]}
&\simeq& -(0.109 \pm 0.015) \ 
{3 - \cos^2\alpha \over 3 + \cos^2\alpha} \label{rhob} \\
\langle \rho^{\,}_{1,-1}(D^{*\pm,0}) \rangle_{[\alpha, \pi-\alpha]}
&\simeq& -(0.216 \pm 0.007) \ 
{3 - \cos^2\alpha \over 3 + \cos^2\alpha} \label{rhod} \\
\langle \rho^{\,}_{1,-1}(\phi) \rangle_{[\alpha, \pi-\alpha]}
&\simeq& -(0.078 \pm 0.014) \ 
{3 - \cos^2\alpha \over 3 + \cos^2\alpha} \label{rhop} \\
\langle \rho^{\,}_{1,-1}(\rho^{\pm,0}) \rangle_{[\alpha, \pi-\alpha]}
&\simeq& -0.265 \ [1- \rho^{\,}_{0,0}(\rho)] \ 
{3 - \cos^2\alpha \over 3 + \cos^2\alpha} \label{rhor} \\
\langle \rho^{\,}_{1,-1}(K^{*\pm}) \rangle_{[\alpha, \pi-\alpha]}
&\simeq& -0.265 \ [1- \rho^{\,}_{0,0}(K^*)] \ 
{3 - \cos^2\alpha \over 3 + \cos^2\alpha} \label{kstarpm} \\
\langle \rho^{\,}_{1,-1}(K^{*0}) \rangle_{[\alpha, \pi-\alpha]}
&\simeq& -0.170 \ [1- \rho^{\,}_{0,0}(K^*)] \ 
{3 - \cos^2\alpha \over 3 + \cos^2\alpha} \,\cdot \label{kstar0} 
\earr
We have explicitely checked that the full expression (\ref{rho+--+})
yields almost identical results [and a negligible imaginary part].

\vskip 12pt
\noindent
{\bf 4 - Comments and conclusions} 
\vskip 12pt

We have computed, within a general factorization scheme, the off-diagonal
helicity density matrix element $\rho_{1,-1}$ of vector mesons produced 
in $e^- e^+ \to \qq \to V+X$ annihilation processes; such element can be
- and in few cases has been - measured via the angular distribution of  
two body decays of the meson in its helicity rest frame. Our results
hold for small $p_T$ and large $z$ hadrons, in particular we expect 
them to hold for heavy mesons which should more easily satisfy such 
requirements. 

Our results for $\phi$, Eq. (\ref{rhop}), are in agreement with data,
Re$\rho^{\,}_{1,-1}(\phi) = -0.11 \pm 0.07$ \cite{opal}; notice that such 
data refer to values of $z > 0.7$ and $\cos\alpha = 0.9$, but still have 
large errors. Our results for $D^*$, Eq. (\ref{rhod}), have the same negative 
sign, but are larger in magnitude than the value found by the OPAL 
collaboration, Re$\rho^{\,}_{1,-1}(D^*) = -0.039 \pm 0.016$ \cite{opal}.
There are good reasons for that: data on $D^*$ are collected for $z > 0.5$, 
and might still contain events to which our calculations do not apply
and for which one expects $\rho^{\,}_{1,-1}=0$; 
one should also not forget that our predictions are somewhat lessened (in 
magnitude) by contributions from $S_{_X}=1$ 
[see comment after Eq. (\ref{rho1-1ts})].

We notice that while the mere fact that $\rho_{1,-1}$ differs from zero is
due to a coherent fragmentation of the $\qq$ pair, the actual numerical
values depend on the Standard Model coupling constants; for example,
$\rho_{1,-1}$ would be positive at smaller energies, at which the one gamma
exchange dominates, while it is negative at LEP energy where the one $Z$
exchange dominates. $\rho_{1,-1}$ has also a peculiar dependence on 
the meson production angle, being small at small and large angles
and maximum at $\theta = \pi/2$.

Such coherent effects in the fragmentation of quarks might not play 
a role in unpolarized observables, where they are usually neglected;
however, they should be taken into account when dealing with more
subtle quantities like off-diagonal spin density matrix elements.
Many of these effects vanish in the limit of small intrinsic momentum
of the hadron inside the jet, $p_T/E_h \to 0$; this happens, for example, 
in the fragmentation of quarks into spin 1/2 hadrons \cite{aamr}. The 
quantity considered here, instead, survives also in the small $p_T$ limit;
we actually exploit such a limit in order to make numerical predictions.

The recent data \cite{opal} are encouraging; it would
be interesting to have more and more detailed data, possibly with 
a selection of final hadrons with the required features for our 
results to hold. A measurement of the $p_T$ of final hadrons and a 
study of the dependence of several observables on its value would
offer many more possibilities of testing both the dynamics of the 
fragmentation process and unusual aspects of the basic interactions;
a measurement of $\rho_{+-}(\Lambda)$ with a selection of $\Lambda$ 
particles with $p_T \not= 0$ is already available \cite{aleph} and in 
agreement with our expectations.


\end{document}